\documentclass[12pt, final]{l4dc2021}
\usepackage{graphicx}
\usepackage{caption}
\usepackage{mathtools} 

\usepackage{dsfont}
\usepackage{xcolor}
\usepackage{algorithm}
\usepackage{algpseudocode} 
\algdef{SE}[DOWHILE]{Do}{doWhile}{\algorithmicdo}[1]{\algorithmicwhile\ #1}%

\title[On exploration requirements for learning safety constraints]{On exploration requirements for learning safety constraints}
\usepackage{times} 
\author{\Name{Pierre-François {Massiani}}\textsuperscript{1, 2} \Email{massiani@dsme.rwth-aachen.de}\\
\Name{Steve {Heim}}\textsuperscript{1} \Email{heim.steve@gmail.com}\\
\Name{Sebastian {Trimpe}}\textsuperscript{1, 2} \Email{trimpe@dsme.rwth-aachen.de}\\
\addr \textsuperscript{1} Intelligent Control Systems Group, Max Planck Institute for Intelligent Systems, Stuttgart, Germany\\
\addr \textsuperscript{2} Data Science in Mechanical Engineering, RWTH Aachen University, Aachen, Germany
}

\renewcommand{\S}{\ensuremath{\mathcal{S}}}
\newcommand{\A}{\ensuremath{\mathcal{A}}}
\newcommand{\Q}{\ensuremath{\mathcal{Q}}}
\newcommand{\SV}{\ensuremath{\S_{\text{V}}}}
\newcommand{\QV}{\ensuremath{\Q_{\text{V}}}}
\newcommand{\Qcrit}{\ensuremath{\Q_\text{crit}}}
\newcommand{\SF}{\ensuremath{\S_{\text{F}}}}

\newcommand{\pichat}{\ensuremath{\pi_{\Kopthat}}}

\newcommand{\Kopt}{\ensuremath{\mathcal{K}}}
\newcommand{\Kopthat}{\ensuremath{\hat{\mathcal{K}}}}
\newcommand{\Constraint}{\ensuremath{\Q_{\text{PI}}}}

\newcommand{\Nominal}{\ensuremath{\pi}}
\newcommand{\OPT}{\ensuremath{\mathcal{OPT}}}

\DeclareMathOperator*{\argmin}{argmin}

\DeclareMathOperator{\cover}{\Pi_\S}
\DeclareMathOperator{\find}{find}
\newcommand{\norm}[1]{\left\lVert#1\right\rVert}

\newcommand{\cost}{\ensuremath{J}}

\begin{document}

\maketitle

\begin{abstract}%
Enforcing safety for dynamical systems is challenging, since it requires constraint satisfaction along trajectory predictions.
Equivalent control constraints can be computed in the form of sets that enforce positive invariance, and can thus guarantee safety in feedback controllers without predictions.
However, these constraints are cumbersome to compute from models, and it is not yet well established how to infer constraints from data.
In this paper, we shed light on the key objects involved in learning control constraints from data in a model-free setting.
In particular, we discuss the family of constraints that enforce safety in the context of a nominal control policy, and expose that these constraints do not need to be accurate everywhere. They only need to correctly exclude a subset of the state-actions that would cause failure, which we call the critical set.
\end{abstract}


\section{Introduction}
Safety is an important and challenging consideration in controller design, especially when deploying systems whose dynamics or environment is difficult to model or otherwise uncertain.
For such systems, data-driven or `learning' based controller design, such as reinforcement learning, has recently had a lot of success~\citep{RLtour2019recht,handful2020mouret,ibarz2021train}.
However, most work has focused on learning optimal controllers, while learning safety is not as well understood. \par
There are various formalizations to ensure safety~\citep{garcia2015comprehensive}. We will consider safety as state constraint satisfaction: there is a \emph{failure set} of states that the system must never visit.
For dynamical systems, however, it is insufficient to simply constrain the controller to stay in the complement of this set: there may be states that are not in the failure set, but from which failure can no longer be avoided within finite time.
Instead, a prediction of unknown horizon length needs to be made to guarantee that the trajectory never passes through the failure set. 
Model-predictive control (MPC) provides a natural description of this problem by predicting these trajectories at run-time. This approach has been successful in many domains, but also has a few drawbacks: in particular, it requires a sufficiently accurate model and substantial computational resources, such that predictions and optimization can be run fast enough during run-time~\citep{hewing2019mpcreview,nubert2020safeMPC}.
\par
We will focus on another route, which provides control constraints that can be applied to a feedback controller, and relies on (pre-)computing controlled invariant sets: a set of states for which there exists a control policy that keeps the system in this set.
These sets can often be computed using tools from the fields of viability theory~\citep{aubin2011viability}, back-reachability~\citep{bansal2017hamilton}, and control barrier functions~\citep{Ames2019CBFthapp}.
However, these tools are also contingent on accurate models and are often too computationally demanding even for offline analysis. \par
The requirement for predictions (and therefore a model) can be entirely sidestepped by directly learning control constraints from data.
The objects of interest here, \emph{control constraints}, are sets in state-action space that can be used to constrain a control policy such that the system never leaves the maximal control invariant set, also called the viability kernel.
Recent work in this direction has provided ways to learn the largest such control constraint~\citep{heim2019learnable, fisac2019reachRL}.
While they successfully sidestep the need for a model, learning the entire maximal control constraint is highly sample-inefficient and often unnecessary. \par
In this paper, we first systematically identify the family of constraints that can guarantee positive invariance, and therefore safety, in the general setting.
We then reformulate the problem of constraining a given nominal policy as a standard optimization problem, which exposes a new object we call the \emph{critical set} of state-action pairs. We prove that the inclusion of the true optimal policy and exclusion of the critical set are sufficient and necessary conditions for a constraint set to enforce safety without resulting in a suboptimal policy.
A key insight is that, because the critical set is a subset of all failure-inducing state-actions, it is not necessary to fully explore the state-action space to converge to a safe, optimal policy.
We believe this insight will inform the design of learning algorithms, in the context of a nominal policy, in two ways: first, it suggests that safe sets do not always need to be estimated conservatively. Second, it shows that a greedy exploration scheme is sufficient for learning safety.
We also provide code to reproduce numerical results in Section~\ref{sec:results}.

\subsection{Related Work}

Model predictive control (MPC) is perhaps the most popular control paradigm for guaranteeing constraint satisfaction~\citep{hewing2019mpcreview}.
Since constraints are enforced along an entire trajectory by formulating an optimization problem along a predicted trajectory, its performance hinges on identifying or learning an accurate model.
In addition, the problem is often formulated with simplified dynamics and quadratic costs, such that it can be efficiently solved via convex optimization.
In contrast, we aim to learn control constraints that can be directly evaluated in a feedback controller.
Learning feedback constraints has substantial challenges; however, once learned, it benefits from being agnostic to the cost formulation and bypasses the need for a model.
It also has much lower computational requirements during run-time.
We see these advantages as particularly relevant in combination with model-free learning algorithms such as deep Q-learning~\citep{xie2020learning}. \par

These control constraints can be computed as sets using tools from back-reachability~\citep{fisac2018general,kaynama2012computing} and viability theory~\citep{wieber2008viability,aubin2011viability,liniger2017real}, or as the superlevel-set of a control barrier function~\citep{Ames2019CBFthapp}.
Computing safe sets for arbitrary dynamical systems, however, relies on computation-intensive, model-based algorithms, which often limits their use to offline analysis, and/or for low-dimensional systems where accurate models are available.
The computational challenge is often addressed by using less accurate, low-dimensional models to obtain set approximations, then bounding the tracking error of the full system~\citep{fridovich2018planning,fridovich2019icra,khadiv2020walking,zhou2020general}.
The model sensitivity challenge is often addressed by refining a set approximation with data, potentially gathered during run-time.
Samples can be used to directly improve the dynamics or disturbance models~\citep{fisac2018general}, but more recently, there has been interest in completely bypassing the model and directly updating the control constraint estimate~\citep{shih2020safesets,fisac2019reachRL,heim2019learnable, rakovic2017reachability,boffi2020certificates, robey2020cbfDemos}.
In this case, the computational cost is typically traded for sample-inefficiency. 
Regardless of the approach, it is common practice to use conservative inner approximations.
This is very reasonable, as only a conservative approximation of the set will maintain its guarantees.
The key insight of our work is that, when used in the context of a specific controller, this conservativeness is unnecessary.
We show that it is sufficient to converge to a family of constraint sets that includes not only (most) conservative approximations, but also many over-approximations, as long as they satisfy specific requirements.
This relaxes the requirements on constraint-learning algorithms, and also shows that greedy, on-policy exploration naturally fulfills these requirements. \par
\section{Viability, Positive Invariance, and Control Constraints} \label{sec:viability}

We consider a discrete-time system with state $s_k \in \S\subset\mathbb{R}^n$, control input $a_k \in \A\subset\mathbb{R}^m$, and dynamics $s_{k+1} = T(s_k, a_k)$, where $k\in\mathbb{N}$ indicates the time-step.
Assumptions on $T$ are introduced in Section~\ref{sec:constrained controller}.
In addition, we consider a set of failure states $\SF$; we assume this to be an absorbing set, e.g., once the system enters this set, it stays there forever and can no longer take actions. We define $\Q = (\S\setminus\SF) \times \A$ as the state-action space without the failure set. \par
Our goal in this section is to define a set $\Constraint$ in state-action space that can be used to constrain a policy such that it induces a positive invariant set in state space.
By constrain, we mean that the policy must map any state $s$ to an action $a$ such that $(s, a) \in \Constraint$~whenever such an action exists.
By positive invariance, we mean that all trajectories that start in a given set in state space will remain in this set forever when following a policy constrained by~\Constraint, and therefore stay out of $\SF$.
\par 
We will start by introducing key concepts related to viability theory and positive invariance, which will allow us to reason about dynamical systems in terms of sets in state-action space without explicitly considering trajectories.
These objects will allow us to reformulate the safe control problem as a standard constrained optimization problem in Section~\ref{sec:reformulation}. \par
\begin{figure}[bth]
    \centering
    \includegraphics[width=0.8\textwidth]{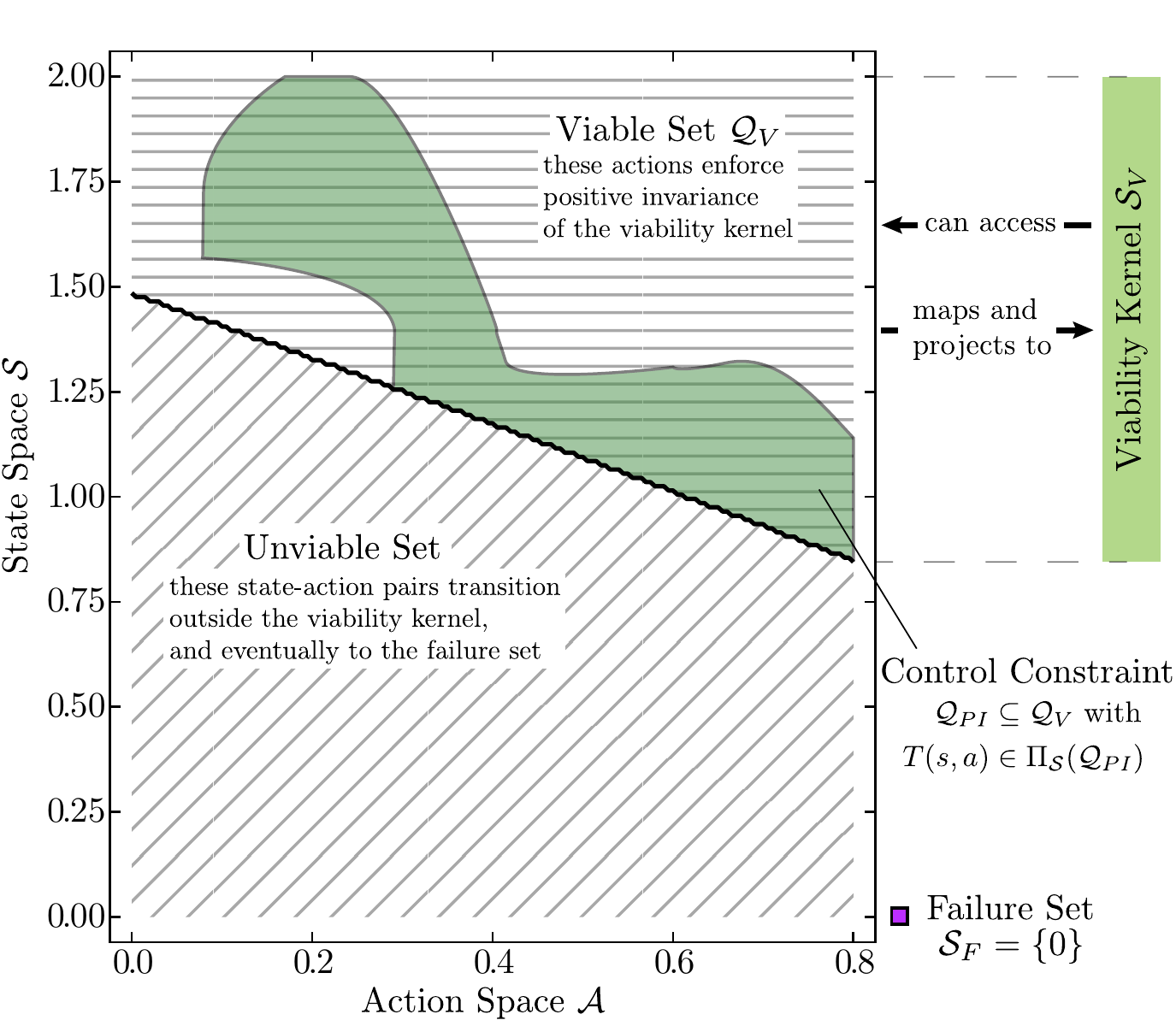}
    \caption{A system's long-term behavior can be captured by the viable set $\QV$ (horizontal hatch) and its complement, the unviable set (diagonal hatch). All state-action pairs in the unviable set form trajectories that terminate in a failure set in finite time (purple). State-action pairs in $\QV$ transition to the viability kernel $\SV$ (green, right). Since $\SV$ is the projection of $\QV$, states in $\SV$ can recursively choose actions in $\QV$ to avoid failure forever. The same holds for any control constraint (green, horizontal hatch) that also has $\SV$ as a projection.}
    \label{fig:annotated}
\end{figure}

\paragraph{Running example} We illustrate these objects on the hovership toy model used in~\citep{heim2019learnable}. This system has a one-dimensional state space $s \in \left[0, 2\right]$ and a one-dimensional action space $a \in \left[0, 0.8 \right]$, and continuous-time dynamics $\dot{s} = a - 0.1 - \tanh (0.75 s)$. We treat this as a discrete-time system by applying control actions with a zero-order hold for one second, and integrating the continuous-time dynamics. Figure~\ref{fig:annotated} illustrates all the objects introduced in this section using this running example. \par
We start by defining the viability kernel $\SV$:
\begin{definition}[Viability Kernel]
    The\emph{ viability kernel} $\SV\subset \S\setminus \SF$ is the maximal set of states from which there exists an action that keeps the system inside $\SV$ (cf.~\citep[Chapter~1.1]{aubin2011viability}).
\end{definition}
In other words, all states outside of $\SV$ will fail within finite time, regardless of the action chosen.
\par
Next, we define the set of state-action pairs that map into the viability kernel:
\begin{definition}[Viable Set]
    The \emph{viable set} $\QV\subset \Q$ is the maximal set of state-actions $(s, a)$ such that $s_{k+1} = T(s_k, a_k) \in \SV$. (cf.~\cite[Definition 2]{heim2019learnable}).
\end{definition}
The viable set $\QV$ can be directly used as a safety constraint: in fact, it is both necessary and sufficient for a safe policy to exclusively map viable states to the viable set.
It is necessary, since all state-action pairs outside $\QV$ map to states outside $\SV$, leading to the failure set in finite time.
It is also sufficient, since all of $\QV$ maps into $\SV$, and $\QV$ can always be accessed by states in $\SV$.
\par

This second property is a key observation that holds for any control constraint~\Constraint: to enforce positive invariance,
the projection of~\Constraint~into state space, $\cover(\Constraint) = \{s, (s, a) \in \Constraint\}$, must be a cover of the set of states mapped from~\Constraint.
We can now define arbitrary control constraints~\Constraint:

\begin{definition}[Control constraint]
We call a \emph{control constraint} a set~$\Constraint\subset \Q$~of state-action pairs where, for all $(s,a) \in \Constraint$ we have $T(s, a)\in\cover(\Constraint)$.
\end{definition}
Control constraints allow us to consider the state-action space in a static sense, without the need to consider trajectories. This is because they map back into their own projection: given a control constraint~\Constraint, the next state-action pair chosen can (and will) be selected from~\Constraint.

\begin{remark}
If $A$ and $B$ are control constraints, and $C$ is any subset of $\Q$, the following hold:
\begin{enumerate}
    \item $A\cup B$ is a control constraint;
    \item If~$\cover(A\setminus C) = \cover(A)$, then $A\setminus C$ is a control constraint;
    \item $A\subseteq \QV$.
\end{enumerate}
\end{remark}
We see that $\QV$ is the largest control constraint, but not every subset of $\QV$ is a control constraint.
Control constraints can be used to guarantee safety for any policy $\pi$: by enforcing $(s, \pi(s))\in\Constraint$, we have a policy that is safe for any initial conditions starting in the projection of~\Constraint.
In the next Section we will show that, if we are only interested in constraining a \emph{specific} policy, we can consider a class of sets with less restrictive requirements.
\section{Constraints Don't Need to be Accurate Everywhere} \label{sec:reformulation}
We now consider the problem of learning constraints for a given nominal policy~\Nominal.
We will show that, by formulating the control policy as a constrained optimization problem, we can consider a much less restrictive set of constraints $\Kopt$ than control constraints.
In addition, this formulation shows that greedily following the nominal policy naturally prioritizes exploring the relevant portion of the state-action space, and off-policy exploration is not necessary.

\subsection{Problem Statement} \label{sec:constrained controller}
In addition to the setting introduced in Section~\ref{sec:viability}, we consider a given nominal policy~\Nominal, and a cost function~$\cost:\Q\to\mathbb{R}$.
We can then define a controller constrained by a set $\Kopt$ as the solution of the optimization problem \OPT:
\begin{equation}
    \begin{aligned}
        \OPT(\Kopt)~:&~s\mapsto~\argmin_{a\in \mathcal{A}}~\cost(s, a),\\
        & \quad \text{s.t.  } \quad(s,a) \in \Kopt.
    \end{aligned} \label{eq:opt}
\end{equation}
For simplicity of exposition, we only consider deterministic policies, and use the euclidean distance to $\pi$ as the cost function $\cost(s, a) = \norm{a - \Nominal(s)}^2$, though other costs can be used instead.
We also require that $\Kopt$ is closed to ensure the existence of the $\argmin$.
\par
If we happen to know $\QV$, we can directly plug~\QV~into~\OPT~to find the viable policy that achieves the lowest cost.
Note that, to ensure that $\OPT(\QV)$ has a well-defined solution, we assume that $\QV$ is closed: to satisfy this assumption, the system dynamics $T$ must generate a closed viable set\footnote{For other costs in~\eqref{eq:opt}, we may require other assumptions on $\QV$.}.
\par
Since $\QV$ is generally unknown, our goal is to find the requirements on $\Kopt$, such that it produces the same policy:
\begin{equation}
    \find~{\Kopt},\quad\text{s.t.}\quad\OPT(\Kopt) = \OPT(\QV).
    \label{eq:problem statement}
\end{equation}
Notice how the only control constraints that satisfy this property are the ones that contain $\OPT(\QV)$.
However, there are many sets $\Kopt$ that are not control constraints, nor even subsets of $\QV$.

\subsection{Admissible Constraints~\Kopt}

\begin{figure}[bth]
    \centering
    \includegraphics[width=0.8\textwidth]{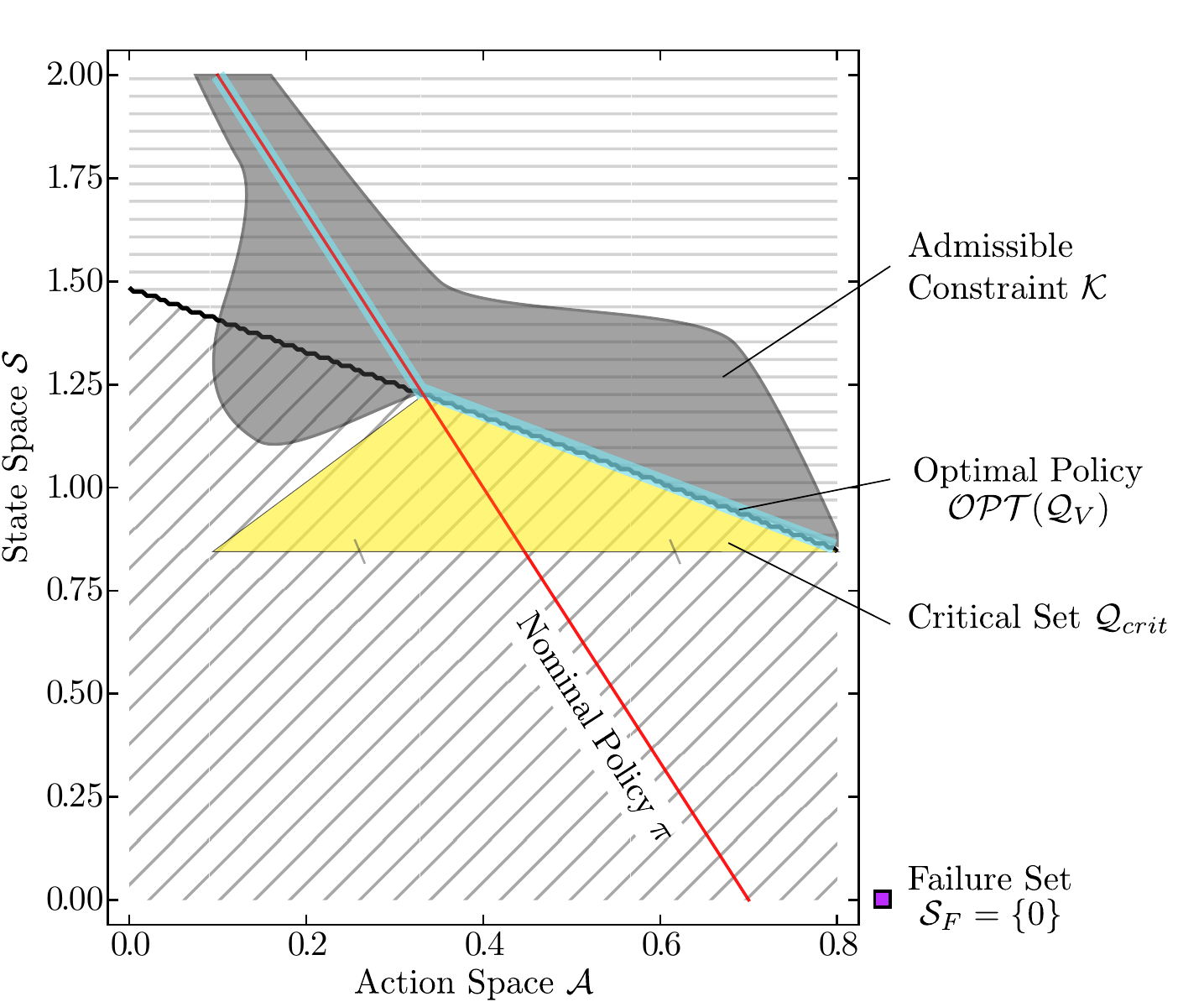}
    \caption{Admissible constraints~\Kopt~(grey), the set~$\OPT(\QV)$~(blue), and the critical set (yellow) are all defined for a nominal policy $\pi$ (red line). The optimal policy is the closest to the nominal one in the viable set (horizontal hatch). The critical set is the set of unviable state-actions (diagonal hatch) closer to $\pi$ than~$\OPT(\QV)$~is. Admissible constraints should include the optimal policy and exclude the critical set. They may contain unviable state-actions that are not critical (grey, diagonal hatch).}
    \label{fig:constraints}
\end{figure}

We have visualized the requirements for~\Kopt~on our running example in Figure~\ref{fig:constraints}.
We start by noting the trivial fact that $\OPT(\QV)$ must be a subset of $\Kopt$ for~\eqref{eq:problem statement} to hold.
This is indeed the only subset of~\QV~that needs to be \emph{included} in~\Kopt, and is highlighted in Figure~\ref{fig:constraints} as the blue shaded area. \par
It is also necessary to \emph{exclude} the set of state-action pairs that achieve lower cost than $\OPT(\QV)$ does, but are not in~\QV; we call this the \emph{critical set}:
\begin{definition}[Critical Set]
    The critical set $\Qcrit$ is the set of all state-actions $(s, a) \in \Q\setminus \QV$ where $s\in\SV$ and $\cost(s, a) \leq \cost(s, \OPT(\QV)(s))$.
\end{definition}
We slightly abuse notations since $J(s, a)$ has the same value for all $a\in\OPT(\QV)(s)$.
This set is highlighted in Figure~\ref{fig:constraints} in yellow.
We conclude that it is both sufficient and necessary for a set~\Kopt~to include $\OPT(\QV)(s)$~and exclude $\Qcrit$ for~\eqref{eq:problem statement} to hold:
\begin{theorem}
Let $\Kopt$ be a closed set that contains $\OPT(\QV)$. Then, $\OPT(\Kopt)(s) = \OPT(\QV)(s)$ for all $s\in\SV$~if, and only if, $\Kopt \cap \Qcrit = \emptyset$. We call such a $\Kopt$ an \emph{admissible constraint}.
\label{theorem:admissible}
\end{theorem}
The proof for this theorem is in the Appendix~\ref{sec:appendix proof}.

\subsection{Greed is Good}
Suppose we wish to learn an estimate~\Kopthat~of an admissible constraint from data with the greedy exploration strategy~\pichat:
\begin{equation}
    \pichat(s) = \begin{cases}
        \OPT(\Kopthat)(s) & \text{if feasible}, \\
        \text{anything} & \text{otherwise}.
    \end{cases}
\end{equation}
In the absence of any domain knowledge, a reasonable choice when $\OPT(\Kopthat)$ is infeasible is to simply follow $\Nominal$.
In fact, following this exploration strategy naturally ensures that $\Qcrit$ will continue to be explored until the learning agent has excluded it.
Unlike in classical reinforcement learning, the exploration requirements are fulfilled by a greedy policy.
\section{Results} \label{sec:results}

As a proof of concept, we demonstrate an on-policy exploration strategy combined with a generic constraint learning algorithm with the structure shown in Algorithm~\ref{alg:explore} to learn an admissible constraint.
Specifically, we use the algorithm developed in~\citep{heim2019learnable}, which models the constraint set with a Gaussian process, and is guaranteed to learn the entire viable set $\QV$ under the assumptions of an optimistic initialization and infinite sampling of $\QV$.
However, instead of minimizing variance, the exploration is driven by following the exploration policy~\pichat.
We test two nominal policies: the linear policy used in the running example in all preceding figures, and a stochastic policy that samples from a uniform distribution of actions.
These two cases can be viewed as the extremes of an $\epsilon$-greedy exploration strategy, with $\epsilon$ set to 0 and 1, respectively.
The Python code to reproduce these results is available at: \url{https://github.com/Data-Science-in-Mechanical-Engineering/edge}



\begin{algorithm}
	\caption{Constraint Set Learner with our Exploration}\label{alg:explore}
	\begin{algorithmic}[1]
		\State \textbf{Input:} Initial state $s_0$, nominal policy $\Nominal$, initial constraint estimate $\Kopthat$
        \Do
            \color{blue}
			\State apply $\pichat(s_i)$:
			\Comment{\parbox[t]{0.4\linewidth}{Greedy On-policy exploration}}
\State \textbf{if} feasible
			    \State \hspace{5mm} $a_i \gets \OPT(\Kopthat)(s_i)$
			\State \textbf{else}
			   \State \hspace{5mm} $a_i \gets \text{anything}$
			\color{black}
			\State $s_{i+1} \gets T(s_i, a_i)$
			\Comment{\parbox[t]{0.4\linewidth}{Sample system dynamics}}
			\State update the constraint set $\Kopthat$ 
			\Comment{\parbox[t]{0.4\linewidth}{Algorithm specific}}
            \State \textbf{if} failed
                \State \hspace{5mm} re-initialize state
 			\doWhile{terminal condition} \Comment{\parbox[t]{0.4\linewidth}{Algorithm specific}}
	\end{algorithmic}
\end{algorithm}


\subsection{Linear Nominal Controller} \label{subsec:linear}
First, we will use the same nominal affine policy in the examples in the preceding sections, with $a=0.7-0.3s$.
We choose a rather conservative initialization of~\Kopthat~to represent a typical situation in which the nominal policy is designed to be reliable around a given operating point. \par
We run our algorithm for two batches of ten episodes each, and at the end of each batch we update the hyper-parameters of the constraint learning algorithm.
At the beginning of each episode, the agent is initialized in a random state that is in the state space projection of the current~\Kopthat. If the agent ever enters the failure set, the episode is immediately terminated, and the last sample is assigned a value of $0$. We also terminate episodes after a maximum of $10$ steps. \par
\begin{figure}[tbh]
    \centering
    \includegraphics[width=0.8\textwidth]{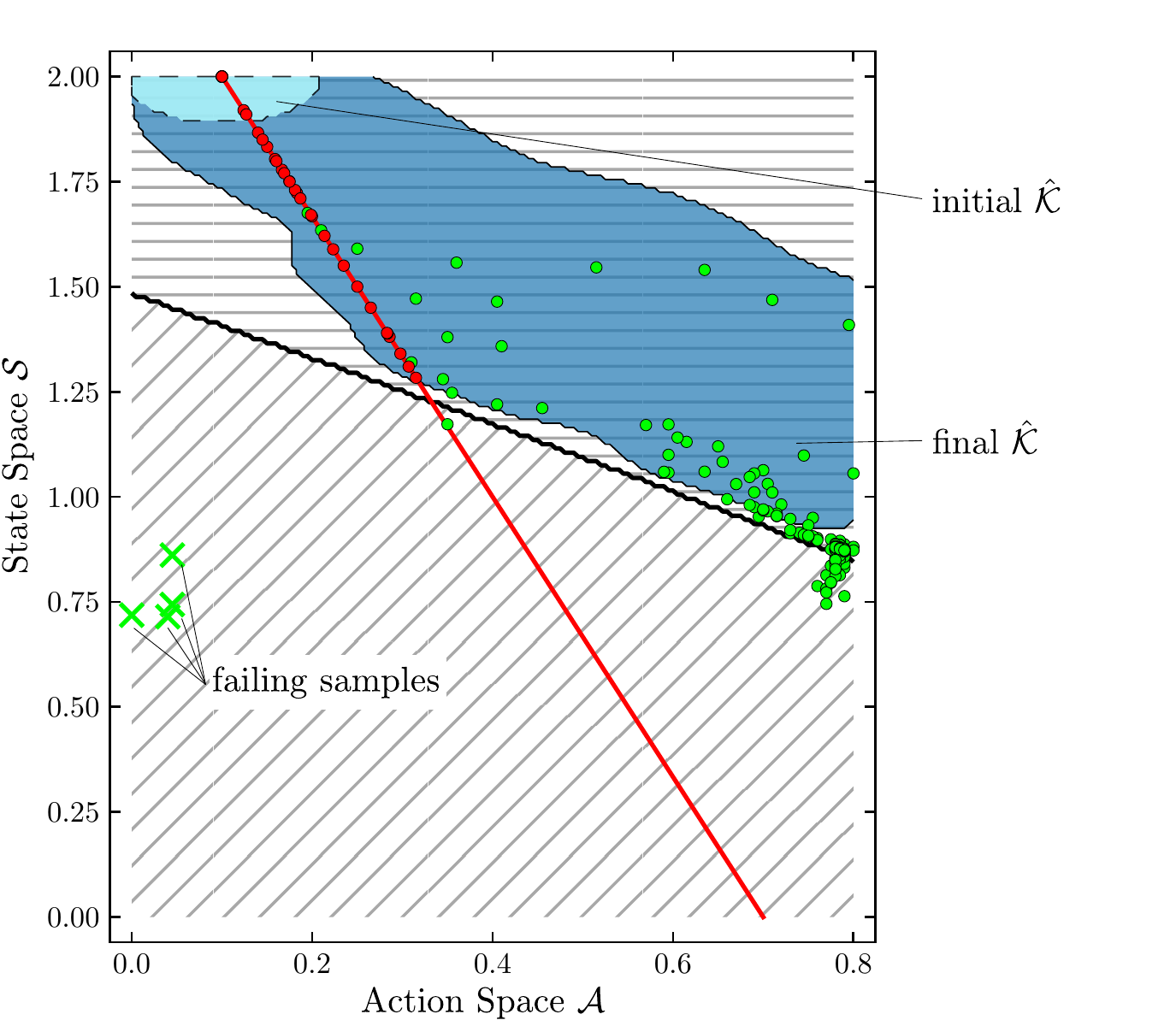}
    \caption{Results for an affine nominal controller (red line) after 20 episodes. The constraint~\Kopthat~grows from a cautious initialization (light blue) to an admissible constraint (dark blue). The red dots are steps at which the constraint allowed the nominal controller, and green dots and crosses are the others. The accumulation of samples near the optimal policy show that~\pichat~converges to it. As expected, the constraint~\Kopthat~does not recover the whole viable set.}
    \label{fig:linear}
\end{figure}

In the end, the agent has collected 196 samples and has learned a safe control policy~\pichat~that ensures safety for nearly the entire viability kernel. The initial and final estimates of $\Kopthat$ are shown in Figure~\ref{fig:linear}.


During the training period, the agent visited the failure set only four times, the last one occurring in episode 15.
The learned policy~\pichat~differs from the lowest-cost policy by at most\footnote{Expressed as percentages of the amplitude of the action set} $10\%$, and on average $2\%$, even though~\Kopthat~underestimates~\QV~by $43\%$.
This example highlights the advantage of our formulation, as we only explore and learn accurate constraints where necessary.

\subsection{Uniformly Random Nominal Controller} \label{subsec:results random}

\begin{figure}[tbh]
    \centering
    \includegraphics[width=0.8\textwidth]{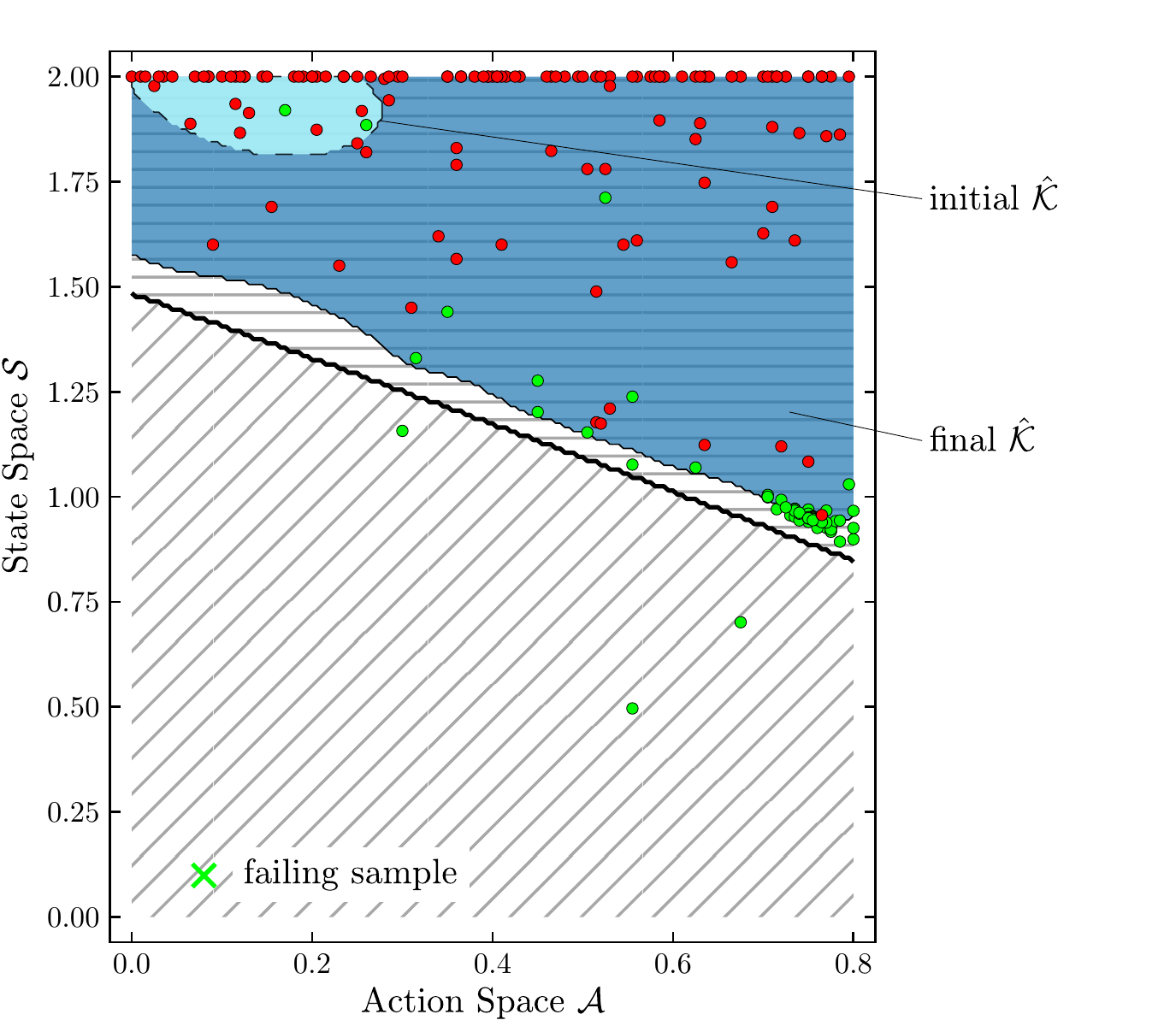}
    \caption{Results for a random nominal controller after 20 episodes. The legend is the same as in Figure~\ref{fig:linear}. Here as well, the constraint~\Kopthat~grows to an admissible constraint. Here, the whole viable set is important for the nominal controller, so~\Kopthat~recovers it.}
    \label{fig:random}
\end{figure}

Next, we set the nominal policy to sample actions from a uniformly random distribution of all actions, while again running our learning algorithm for two batches of ten episodes and with the same termination criteria.
In this case, the critical set contains all state-actions that transition outside of the viability kernel, and the policy continues to push the constraint estimate outwards. Unsurprisingly, $\Kopthat$ converges towards the true viable set, as can be seen in Figure~\ref{fig:random}.
Here, the discrepancy between~\pichat~and the lowest-cost policy~is equivalent to how much~\Kopthat~underestimates~\QV, since the nominal policy would sample from all of~\QV. In our experiment, this reaches $9\%$ after 197 samples.
This example highlights the applicability of our exploration strategy to stochastic policies. 
\section{Discussion}

We have presented a formulation of constrained controllers as sets in state-action space, which allows classes of constraints to be easily defined and analyzed in a static sense, as positive invariance reduces to considering sets and their projection into state space.
With this formulation, we identify the \emph{viable set} as the set of all safe state actions, and \emph{control constraints} as all possible subsets that induce positive invariance, and thus guarantee safety over their projection into state space.
Finally, we identify \emph{admissible constraints} as sets used to safely constrain a nominal policy, and show that these are a separate family from control constraints: admissible constraints need not be control constraints.
Our key insight is that admissible constraints only need to accurately exclude a \emph{critical set} of state-actions, and that a policy greedily following the nominal policy naturally prioritizes exploring this set.
This property also suggests that, when estimating safe sets, it is not necessary to restrict algorithms to conservative estimates.
\par
Our long-term goal is to learn safe control policies, and not just safety constraints for stationary policies.
However, how changing the nominal policy affects the minimum sampling requirements needs to be clarified.
Although the arguments we have made hold for each instance of a moving policy, the critical set will move together with the policy.
We expect that the insights developed in this paper can nonetheless be used to guide exploration and improve sample efficiency.
Our focus in this paper has been on a purely data-driven scenario.
As is often the case, we also expect that leveraging model-based predictions can further improve sample efficiency and will be critical to scaling up to real-world applications.
A straightforward approach is to plan over trajectories~\citep{buisson2020actively, hewing2019mpcreview}.
A different approach we find promising is to learn parsimonious models based on the required properties of a control constraint that we have elucidated in Section~\ref{sec:viability}: we do not need models that make accurate state predictions, but only predict if sets map into their projection.

\appendix
\section{Proof of Theorem~\ref{theorem:admissible}}
\label{sec:appendix proof}
We first introduce the \emph{action slice} of $\Kopt$ in state $s$:
\begin{equation*}
    \Kopt[s]= \left\{a\in\mathcal{A}, (s, a) \in \Kopt\right\}.
\end{equation*}
\begin{proof}
First, suppose that for all $s\in\SV,~\OPT(\Kopt)(s) = \OPT(\QV)(s)$. We prove that any element $(s, a)$ of $\Qcrit$ is not in $\Kopt$. Since $(s, a)$ is not in $\QV$, we have:
\begin{equation*}
    a \notin \argmin_{b\in\Kopt[s]}J(s, b).
\end{equation*}
Yet, by definition of $\Qcrit$ as the set of unviable state-actions that achieve lower $J$ than $\OPT(\QV)$, we also have:
\begin{equation*}
    J(s, a) \leq \min_{b\in\Kopt[s]}J(s, b).
\end{equation*}
Combining the last two equations immediately shows that $a \notin \Kopt[s]$, which proves the implication.\par
Conversely, we prove that $\OPT(\Kopt)(s) = \OPT(\QV)(s)$ for $s\in\SV$. We proceed by proving the double inclusion. Let $a$ in $\OPT(\Kopt)(s)$.
We immediately have:
\begin{equation}
    J(s, a) \leq J(s, \OPT(\QV)(s)),\label{eq:Ja lower Jpistar}
\end{equation}
since $\OPT(\QV)$ is contained in $\Kopt$. Thus, $a$ is at least as close to $\pi(s)$ than $\OPT(\QV)(s)$. Therefore, $(s, a)$ is either in \Qcrit~or in \QV. Since $\Kopt\cap\Qcrit=\emptyset$, then $(s, a)$ is in $\QV$, and~\eqref{eq:Ja lower Jpistar} is an equality. This means that $a\in \OPT(\QV)(s)$, and shows the first inclusion.\par
Now consider $a$ in $\OPT(\QV)(s)$. We immediately have $(s, a)\in\Kopt$, since it contains $\OPT(\QV)$. Since $a$ is a closest viable actions to $\pi(s)$ and $\QV$ contains $\OPT(\Kopt)(s)$, we also have $J(s,a) = J(s,\OPT(\Kopt)(s))$. This shows $\OPT(\QV)(s)\subset\OPT(\Kopt)(s)$, and concludes the proof.
\end{proof}


\acks{We thank Friedrich Solowjow and Alexander von Rohr for frequent insightful discussions. This work was funded, in part, by the Cyber Valley Initiative.}

\bibliography{bibliography}

\end{document}